\documentclass[pra,aps,groupedaddress,floatfix,twocolumn,superscriptaddress,showpacs]{revtex4-1}
\usepackage[utf8]{inputenc}
\usepackage[T1]{fontenc}
\usepackage{amsmath}
\usepackage{amssymb}
\usepackage{epsfig}
\usepackage{bm}
\usepackage{pifont}
\usepackage{color}
\usepackage{graphicx}
\usepackage{epstopdf}
\usepackage{bm}
\usepackage{tabularx}

\newcommand{\beq}{\begin{equation}}
\newcommand{\eeq}{\end{equation}}
\newcommand{\bea}{\begin{eqnarray}}
\newcommand{\eea}{\end{eqnarray}}
\newcommand{\f}[2]{\frac{#1}{#2}}
\newcommand{\pf}[2]{\frac{\partial #1}{\partial #2}}

\newcommand{\la}{\langle}
\newcommand{\ra}{\rangle}

\newcommand{\mD}{\mathcal D}
\newcommand{\mZ}{\mathcal Z}
\newcommand{\mU}{\mathcal U}
\newcommand{\tn}{\textnormal}

\newcommand{\dm}{\delta m}
\newcommand{\mbar}{\bar m}
\newcommand{\kf}{k_\textnormal{F}}
\newcommand{\ef}{\epsilon_\textnormal{F}}
\newcommand{\efg}{E_{\textnormal{FG}}}
\newcommand{\bef}{\beta\ef}
\newcommand{\qpsi}[1]{\hat{\psi}_{#1}(x)}
\newcommand{\qpsidagger}[1]{\hat{\psi}_{#1}^{\dagger}(x)}
\newcommand{\pderiv}[2]{\frac{\partial #1}{\partial #2}}

\begin{document}

\title{Surmounting the sign problem in nonrelativistic calculations:\\
A case study with mass-imbalanced fermions}

\author{Lukas Rammelm\"uller}
\affiliation{Institut f\"ur Kernphysik (Theoriezentrum), Technische Universit\"at Darmstadt,
D-64289 Darmstadt, Germany}

\author{William J. Porter}
\affiliation{Department of Physics and Astronomy, University of North Carolina, Chapel Hill, NC, 27599, USA}

\author{Joaqu\'in E. Drut}
\affiliation{Department of Physics and Astronomy, University of North Carolina, Chapel Hill, NC, 27599, USA}

\author{Jens Braun}
\affiliation{Institut f\"ur Kernphysik (Theoriezentrum), Technische Universit\"at Darmstadt,
D-64289 Darmstadt, Germany}
\affiliation{ExtreMe Matter Institute EMMI, GSI, Planckstra{\ss}e 1, D-64291 Darmstadt, Germany}

\begin{abstract}
	The calculation of the ground state and thermodynamics of mass-imbalanced Fermi systems is a challenging many-body problem.
	Even in one spatial dimension, analytic solutions are limited to special configurations and numerical progress with standard
	Monte Carlo approaches is hindered by the sign problem. {The focus of the present work
		is on the further development of methods to study imbalanced systems in a fully nonperturbative fashion.}
	We report our calculations of the ground-state energy
	of mass-imbalanced fermions using two different approaches {which are also very popular in the context of the theory of the
	strong interaction (quantum chromodynamics, QCD)}: (a) the hybrid Monte Carlo algorithm with imaginary mass imbalance,
	followed by an analytic continuation to the real axis; and (b) the complex Langevin algorithm.
	We cover a range of on-site interaction strengths that includes strongly attractive as well as strongly repulsive {cases which
		we verify with nonperturbative renormalization group methods and perturbation theory.
		{Our findings indicate that, for
		strong repulsive couplings, the energy starts to flatten out, implying}
		interesting consequences for short-range and high-frequency correlation functions. Overall, our results clearly indicate that
		the Complex Langevin approach is very versatile and works very well for {imbalanced Fermi gases with both attractive and repulsive interactions.}}
\end{abstract}

\maketitle

\section{Introduction}
Ultracold quantum gases have given us the opportunity to directly observe many-body physics at work in an unprecedented way.
Over the last few decades, with the advent of laser trapping and cooling techniques, experimentalists have
progressively achieved a previously unimagined {degree of control} for a wide range of atomic systems. {As a consequence, the variety of quantities that can be
	measured with precision has grown dramatically }~\cite{Celi2017}. As is now well known, tuning across magnetic Feshbach
resonances allows for the interaction strength to be varied essentially at will. Additionally, the fine tuning of optical
trapping potentials has enabled the study of lattice models of direct relevance to condensed matter physics~\cite{Lewenstein2012},
including systems in low dimensions (by highly constrained traps). Similarly,
the clever use of the internal nuclear states of alkali and alkali-earth atoms has made it possible to probe systems with multiple
internal degrees of freedom, i.e. with SU($N$) flavor symmetry~\cite{Cazalilla2014, Pagano2014}. Most relevant for this work, experiments involving
atomic mixtures of fermionic (or even fermionic {\it and} bosonic) species has facilitated access to {\it mass-imbalanced} situations, a
case that is interesting for a variety of {reasons but has been studied less than its spin-imbalanced counterpart}.
Recent experimental setups include fermionic mixtures of $^6$Li and $^{40}$K \cite{Trenkwalder2011, Spiegelhalder2009, Tiecke2010}.
{Moreover, other systems with
	mixtures  of  a  variety  of  suitably chosen different  fermion species
	(such as ${}^6$Li, ${}^{40}$K, ${}^{161}$Dy, ${}^{163}$Dy, and ${}^{167}$Er) appear within reach
	in the  future (see, e.g., Refs.~\cite{GrimmPC, Lu2012, Frisch2013}).}

{Given the rapid experimental progress, in particular with respect to spin- and mass-imbalanced quantum gases,
	this work aims at the further theoretical development of stochastic frameworks required for {\it ab initio} studies of such
	systems in any dimension. In other words, with our developments in the present paper,
	we particularly aim at aspects which can currently only be
	accessed in a very limited fashion with conventional Monte Carlo~(MC) methods, if at all. To test our developments, we
	examine the equation of state of mass-imbalanced Fermi mixtures when confined to one-dimensional (1D)
	situations. Although 1D systems are also experimentally relevant (see, e.g., Ref.~\cite{Zuern2013}), we do not aim at a high-precision
	calculation of the equation of state with our present study.
	Our goals are rather of methodological nature. From the latter standpoint,
	the 1D limit is appealing since the running times of the computations are comparatively short and it is therefore possible to take
	vastly more data than in higher dimensions. This allows to reduce systematic errors (e.g. by studying large lattice sizes) and focus on the underlying methods.
Moreover, as is well known, 1D systems} of fermions with contact interaction are typically solvable by the Bethe ansatz technique for arbitrary
particle numbers \cite{Guan2013}. In some cases the analytic
investigations have been complemented by numerical studies (see e.g. \cite{Rammelmueller2015, Rammelmueller2017, Volosniev2014, Andersen2016}).
{However, as soon as the system involves particles of {general unequal}
	masses, an analytic solution is currently out of reach. While the two-body problem
	can of course be solved, the solution of mass-imbalanced few-body systems is restricted to special mass configurations \cite{Harshman2017, Olshanii2015},
	or infinite interaction strength \cite{Dehkharghani2016}, or specific boson-fermion mixtures~\cite{Volosniev2017}.
	In any case, the existence of analytic solutions in some cases and the absence of theoretical
results in other cases represents a further motivation for the developments discussed in the present work.}

It is worth noting that progress has been made in related cases such as the half-filled {asymmetric} Hubbard model in 1D
\cite{Liu2015} as well as 3D \cite{Philipp2016}, which does not feature a sign problem and is directly connected to the Falicov-Kimball
model in the limit of {large asymmetry}
(and is thus of interest to a sector of the condensed-matter community). Cases away from half filling were also studied in
Refs.~\cite{Farkasovsky2008, Batrouni2009}. Furthermore, exact diagonalization studies have addressed harmonically trapped systems of up to $10$ particles
\cite{Pecak2016, Pecak2017}. While these methods provide results for the few-body regime, it is challenging to extend them beyond low particle numbers {or to higher dimensions} due
to the prohibitive scaling of memory requirements. By trading the precise knowledge of the wavefunction for answers to specific questions, i.e. by the use of
Quantum Monte Carlo methods (QMC) to estimate specific correlation functions, {we may not only push further in particle number but also to higher-dimensional
systems}, which has led to great success
in spin- and mass-balanced systems. As one moves to mass-imbalanced systems, however, further restrictions appear, namely the infamous sign
problem, and computational effort is again prohibitive.

One way to circumvent the exponentially large computational cost at finite mass difference is to
``take a detour via the complex plane". Motivated by the use
of an imaginary chemical potential in the investigation of the QCD phase diagram \cite{deForcrand2002}, it was recently shown that the same idea also
is useful for nonrelativistic fermions in 3D \cite{Braun2013} and 1D \cite{Loheac2015} at finite temperature. A natural move from imaginary polarization is the step to imaginary mass imbalance, which was investigated in Ref.~\cite{Roscher2014} and later applied to the ground state of a unitary Fermi gas on the
lattice \cite{Braun2015}. Another approach to circumvent the sign problem by using complex numbers is
the {so-called complex Langevin (CL) method, which} is
a complex generalization of the idea of stochastic quantization. {Considerable progress was made in the last decade in understanding when that
	method is valid and useful for relativistic theories (see e.g. Refs.~\cite{Aarts2009, Aarts2010, Aarts2010a, Aarts2011}) as well as for nonrelativistic systems (see e.g Refs.~\cite{Adami2001, Yamamoto2015, Loheac2017}).}

Below, {we first describe the model underlying our studies,
relevant} scales, and dimensionless parameters, and elaborate on both the imaginary-mass and CL
approaches. After that, we present our results for the ground-state energy and compare the two methods at finite mass imbalance in the case of
attractive {interactions in Sec.~\ref{sec:res}. In the same section,
	our main result, namely the equation of state as a function of both mass imbalance and interaction strength (both attractive
and repulsive) is shown. We summarize and present our conclusions in the last section.}

\section{Model, Scales, and parameters}

In this work we focus on nonrelativistic 1D fermions with contact interaction among the spin species, which are governed by the Hamiltonian
\beq
\hat H = \hat T + \hat V,
\label{Eq:model}
\eeq
{with
	\beq
	\begin{split}
		\hat T &= \sum_{s=\uparrow,\downarrow}{\int{{\rm d}x\ \qpsidagger{s}\left(-\f{\hbar^2}{2 m_s} \partial_x^2\right)\qpsi{s}}},\\
		\hat V &= g\int{{\rm d}x\ \qpsidagger{\uparrow}\,\qpsi{\uparrow}\,\qpsidagger{\downarrow}\,\qpsi{\downarrow}}.
		\label{Eq:modelParts}
	\end{split}
	\eeq
	Here, $\qpsidagger{s}$ and} $\qpsi{s}$ denote operators that create and annihilate fermions of spin $s$, respectively.
{Note the spin dependence of the mass in the kinetic part $\hat T$ renders the model insoluble},
as opposed to the integrable Gaudin-Yang model \cite{Gaudin1967, *Yang1967}.

The above expressions describe dilute Fermi gases when the effective interaction range $r_0$ is much smaller than the average interparticle distance $\sim \kf^{-1}$, with $\kf=\f{\pi}{2}n$
being the {Fermi momentum. In}
such systems, the sole physical parameter describing the interaction between {particles is the
	$s$-wave scattering} length $a$ which is connected to the coupling through $g = 2/a$ (see e.g. Ref. \cite{Barlette2000}).
{In our case, additional} physical input are the total particle
number $N=N_\uparrow + N_\downarrow$ and the box size $L$, which we use to define the conventional dimensionless coupling $\gamma = g/n$, with
$n=N/L$ being the particle density.

{From now on, we work
	in units such that \mbox{$k_\tn{B} = \hbar = 1$} and} normalize our results for the ground-state energy using the energy of the noninteracting
mass- and spin-balanced Fermi gas in the continuum
\beq
\efg = \f{1}{3}N\ef,
\eeq
where $\ef = \kf^2/2$. To simplify the discussion of mass imbalanced systems, we define the dimensionless mass-imbalance {parameter
	\beq
	\mbar = \f{m_\uparrow - m_\downarrow}{m_\uparrow + m_\downarrow}\,,
	\label{Eq:mbar}
	\eeq
	which is consistent with the literature~\cite{Roscher2014,Braun2015}.}
Note that the system is invariant under~$\bar{m}\to -\bar{m}$
as long as the system is unpolarized, which
is always the case in this work.

\section{Many-body methods}\label{sec:mbm}

Here we present the essential ingredients of our ground-state formalism for Fermi gases with short-range interactions described by Eq.~\eqref{Eq:modelParts}.
We start with the approach previously employed to 1D, 2D, and 3D Fermi gases with equal masses and attractive interaction on a lattice with periodic boundary conditions \cite{Rammelmueller2015, Rammelmueller2016, Drut2012}. The partition {sum $\mZ_\beta$ is written as
	\beq
	\mZ_\beta = \la\psi_0|\,e^{-\beta\hat{H}}\,|\psi_0\ra \equiv \la\psi_0|\,\hat{\mU}_\beta\,|\psi_0\ra,
	\label{Eq:partitionFunctionOperator}
	\eeq
	which projects the guess state $|\psi_0\ra$ onto the ground state in the limit $\beta\rightarrow\infty$. Here,~$\beta$ refers
to the extent of the imaginary time direction.}
The central object to compute is the
transfer matrix $\hat{\mU}_\beta$, which is challenging for any nontrivial $\hat H$ because $\hat T$ and $\hat V$ do not commute. To deal with the two-body operator $\hat V$, a discretization of the imaginary time axis is performed followed by symmetric Trotter-Suzuki factorization~\cite{Trotter1959, *Suzuki1976}. This is in turn followed by a Hubbard-Stratonovich (HS) transformation to replace the quadratic occurrence of the density operator with a linear one coupled to an auxiliary field $\sigma (x,\tau)$ (see Refs.~\cite{Stratonovich1957, *Hubbard1959}). {Eventually, these steps yield the
	following path integral:
	\beq
	\mZ_\beta = \int{\mD\sigma \det{ U_\beta^{\uparrow}[\sigma]} \det{ U_\beta^{\downarrow}[\sigma]}}\,.
	\label{Eq:partitionFunction}
	\eeq
The determinants} in the above expression are taken
over the single-particle representation of the respective (HS-transformed) transfer matrices $U_\beta^{s}[\sigma]$, which reflects the use of a Slater determinant as a trial state $|\psi_0\ra$ (see Ref.~\cite{Assaad2008}). It is
crucial for conventional Monte Carlo approaches that the product of these
determinants be non-negative, since only then one may interpret the integration kernel as a probability measure:
\beq
\mZ_\beta \equiv \int \mD \sigma \ P[\sigma] \equiv \int \mD \sigma \ e^{-S[\sigma]}\,.
\label{Eq:pdf}
\eeq
{Here, we defined the action $S[\sigma] = -\ln{P[\sigma]}$, which} is real only when $P[\sigma]$ is positive. To evaluate the path integral, we apply the
\emph{hybrid Monte Carlo} (HMC) algorithm \cite{Duane1987, *Gottlieb1987}, which is an essential method for lattice QCD calculations. The objective of HMC
is to perform global updates on the auxiliary field $\sigma$ as opposed to a number of random local updates. This goal is achieved by introducing a conjugate
momentum field $\pi(x,\tau)$ and multiplying the path integral by an immaterial constant factor:
\beq
\mZ_\beta \equiv \int \mD \pi \int \mD \sigma \ e^{-\mathcal H[\sigma,\pi]}\,,
\label{Eq:pdf2}
\eeq
where
\beq
\mathcal H [\sigma,\pi] \equiv S[\sigma] + \int {\rm d}x \int {\rm d}\tau \; \frac{1}{2}{[\pi(x,\tau)]^2}\,.
\eeq
To obtain an updated field configuration, the equations of motion, given by

\beq
\begin{split}
	\pderiv{\sigma}{t} &= \frac{\delta\mathcal {H}}{\delta\pi} = \pi\,,\\
	\pderiv{\pi}{t} &= -\frac{\delta\mathcal{H}}{\delta\sigma} = -\frac{\delta S[\sigma]}{\delta \sigma}
\end{split}
\label{Eq:HMCmotion}
\eeq

are integrated along a trajectory of length $\sim 1$ in the fictitious HMC time $t$. By this on-shell propagation of the auxiliary-field,
governed by the auxiliary classical Hamiltonian $\mathcal H[\sigma,\pi]$ (whose value is preserved throughout the evolution), the acceptance rate for the
Metropolis accept-reject step is {almost $100\%$, as} the same Hamiltonian $\mathcal H$ is used to decide that step. The latter is allowed
because, as noted above, the introduction of the field $\pi$ into the action is immaterial to the dynamics of the {system.
	Indeed, it factors out of $\mZ_\beta$ completely.}

To calculate physical observables we can take derivatives with respect to
associated source terms introduced in {the action $S[\sigma]$.}
In this work, we focus on the ground-state energy, which we obtain by taking a log-derivative of the partition sum with respect to imaginary time:

\beq
E_\beta = -\pf{\ln \mZ_\beta}{\beta} = -\f{1}{\mZ_\beta}\pf{\mZ_\beta}{\beta}.
\label{Eq:energyEstimator}
\eeq

As mentioned above, for a method based on importance sampling \cite{Metropolis1953}, a positive probability measure $P[\sigma]$ in Eq.~\eqref{Eq:pdf}
needs to be guaranteed, or in other words: the {action $S[\sigma]$ must} be real.
This is only the case for systems with an even number of equally populated spin species with uniform {masses. In this case,
	the} transfer matrices of all spin species are equal. Additionally, within our approach, it is necessary for the interaction to be attractive such that the operator $\hat{\mU}_\beta$ in
Eq.~\eqref{Eq:partitionFunctionOperator} is real. For any other fermionic system our QMC-based approach is subject to the infamous sign problem, rendering the
simulation time exponentially growing in system size (particle number or spatial extent, depending on the specific algorithm). Below, we elaborate on the two
methods used here to circumvent the sign problem for systems of fermions {with unequal masses as well as repulsive interactions.}

\subsection{Imaginary mass-imbalance}

{As outlined in the previous section, it is}
necessary to provide a non-negative integral kernel in Eq.~\eqref{Eq:pdf} to enable QMC sampling. By choosing the masses of the particles to be complex and such that they satisfy the condition $m_\uparrow = m_\downarrow^*$,
one can show that the transfer matrices $U_\beta^{\uparrow}[\sigma]$ and $U_\beta^{\downarrow}[\sigma]$ for spin-up and -down fermions are complex conjugate of each other. It is instructive to write the masses as
\beq
\begin{split}
	m_\uparrow = m_0 + i\f{\dm}{2},\\
	m_\downarrow = m_0 - i\f{\dm}{2},
	\label{Eq:idm}
\end{split}
\eeq
which using Eq.~\eqref{Eq:mbar} yields
\beq
\mbar = i\f{\dm}{2 m_0}.
\eeq
{In the following, we set~$m_0=1$ which fixes the scale for the masses in our calculations.}

With these definitions, the product of the determinants can be written as an absolute square and thus remains positive semidefinite for arbitrary imaginary mass asymmetry:
\beq
P[ \sigma ] = \det U^{\uparrow}_\beta[\sigma] \, \det U^{\downarrow}_\beta[\sigma] = |\det U^{\uparrow}_\beta[\sigma]|^2.
\label{Eq:measure}
\eeq
The partition sum (\ref{Eq:partitionFunction}) can now be obtained via standard QMC methods and we are able to extract observables
as a function of the imaginary mass imbalance $\mbar$. In order to obtain physical results, however, we need to
perform an analytic continuation to real mass imbalance via e.g. a {polynomial fit or a fit to a Pad\'e approximant. Strictly} speaking,
such a continuation to the real plane is only defined if the partition sum $\mZ_\beta$ is an analytic
{function of $\mbar$, a fact that}
is not trivially confirmed in practice. To gain analytic insight, however, we discuss below the noninteracting Fermi gas along with our results using the QMC approach discussed above.

It is important to note here that this approach is fully nonperturbative. The results do contain systematic uncertainties, but those are by definition controllable as they arise from the discretization of spacetime. Naturally, the analytic continuation has
limitations and actually fails at very {high mass imbalances~\cite{Braun2015}, as we will also discuss below. For
low to intermediate mass-imbalances, however, the use of imaginary mass-imbalances enable} the calculation of few- to many-body properties of
Fermi gases in arbitrary dimension. Below, we will use the abbreviation iHMC to refer to the above approach of combining HMC data at imaginary
$\bar m$ followed by analytic continuation.

\subsection{Complex Langevin dynamics}

Instead of adapting the method such that a positive probability measure is guaranteed, one may rethink the update process of the auxiliary field $\sigma$ altogether. More specifically, we may let $\sigma$ evolve according to a different equation of motion
\beq
\pderiv{\sigma(t)}{t} = -\frac{\delta S[\sigma]}{\delta  \sigma} + \eta(t),
\label{Eq:realLangevin}
\eeq
i.e. the \emph{Langevin equation}. Here, $t$ is a ficticious Langevin time and $\eta$ comprises a random noise field with expectation value $\la\eta(t)\ra = 0$
and autocorrelation $\la\eta(t)\eta(t')\ra=2\delta_{t,t'}$. The above expression is borrowed from the context of statistical physics, where it describes the
stochastic movement of a heavy (slow) particle immersed in a rapidly changing background of lighter (fast) particles, i.e. the \emph{Brownian motion}. In a
computational context, on the other hand, the use of Eq.~\eqref{Eq:realLangevin} is termed (real) \emph{Langevin dynamics} (RL), whose foundation lies in the
concept of \emph{stochastic quantization}. The latter interprets the stationary distribution of a stochastic process as the probability measure in the path
integral of the corresponding Euclidian field theory \cite{Parisi1981}.

Although the RL algorithm is again restricted to real actions, complex actions can be considered by complexifying the auxiliary field $\sigma$. We then
obtain a new set of equations of motion \cite{Parisi1983}:
\beq
\begin{split}
	\pderiv{\sigma_R(t)}{t} &= -\tn{Re}\left[\frac{\delta S[\sigma]}{\delta \sigma}\right] + \eta(t)\,, \\
	\pderiv{\sigma_I(t)}{t} &= -\tn{Im}\left[\frac{\delta S[\sigma]}{\delta \sigma}\right]\,.
\end{split}
\label{Eq:complexLangevin}
\eeq
Unfortunately, there is no rigorous mathematical foundation for the CL
approach and despite its elegance there are two caveats with the method. First of all,
convergence is not guaranteed due to numerical instabilities and, even if convergence is
achieved, it is not certain that the correct result is reproduced. The former difficulty is of
numerical nature and has been cured by using adaptive time-step solvers \cite{Aarts2010}.
{The issue regarding the correctness of the result is much}
more delicate and is due to singularities in the imaginary part of the
auxiliary field \cite{Salcedo2016, Aarts2010a, Aarts2011}. More precisely, these singularities
occur in our case through the use of an HS transformation that depends on $\sin \sigma$:
\beq
\sin\sigma = \sin\sigma_R \cosh\sigma_I + i\cos\sigma_R\sinh\sigma_I.
\label{Eq:HSdiverging}
\eeq
Thus, the imaginary direction is not bounded and expectation values of observables must be assumed to be contaminated by singularities i.e. cannot be trusted without further analysis even if convergence is achieved.

To prevent the CL algorithm from uncontrolled ``excursions" in the complex plane, the insertion of a ``regulator" in the equation of motion was proposed recently~\cite{Loheac2017}. The discretized equations of motion then read
\beq
\begin{split}
	\delta{\sigma}_R\ &=\ -\tn{Re}\left[\frac{\delta S[\sigma]}{\delta \sigma}\right]h_t\ -\ 2\xi\sigma_{\tn{R}}h_t + \ \eta \sqrt{h_t}\,,\\
	\delta{\sigma}_I\ &=\ -\tn{Im}\left[\frac{\delta S[\sigma]}{\delta \sigma}\right]h_t\ -\ 2\xi\sigma_{\tn{I}}h_t\,,
\end{split}
\label{Eq:discreteCL}
\eeq
where $h_t$ is the (adaptive) step size in CL time $t$. The parameter $\xi$ determines the strength of the regulating term which can be thought of as a damping force that keeps the auxiliary field from wandering to large values of~$\sigma$. Of course, this term represents a systematic influence whose effect needs to be studied carefully. Practically, we can calculate observables at different values of~$\xi$ and then consider the extrapolation $\xi\rightarrow 0$. We have checked this issue carefully and observe the same convergence pattern as reported in Ref.~\cite{Loheac2017}. In our explicit calculations, we have found
that~$\xi$ has to be chosen such that
the regulator term is rendered sufficiently large compared to the average magnitude of the
drift term~$\sim\delta S/\delta \sigma$.
At the same time, we also have to keep~$\xi$
sufficiently small to ensure that the regulator term does not exceed the average magnitude of the drift term and therefore dominates the physics.

Although
several runs at different values of $\xi$ are needed to obtain results, this procedure only introduces a linear increase of computational
effort as the simulation time of a single run does not depend on the value of $\xi$ explicitly.
To illustrate the extrapolation procedure, we show in Appendix~\ref{App:A} how the energy of a Fermi gas with~$\bar{m}=0.3$ and~$\bar{m}=0.6$ has been extracted from the numerical data obtained with different values of~$\xi$. In addition, a discussion of the role of the parameter~$h_t$ controlling the (adaptive) step size in Eq.~\eqref{Eq:discreteCL} can be found in Appendix~\ref{App:A} as well.

Using Eq.~\eqref{Eq:discreteCL}, it is possible to estimate path integrals that would be subject to a sign-problem in conventional QMC approaches. Thus, we have a method at hand to study, in a fully nonperturbative way, many-body systems of mass- and spin-imbalanced Fermi gases, at least potentially without constraints on any imbalance parameters. In the following, we apply the method to 1D fermions with arbitrary mass imbalance and underline its correct behavior by comparison with other approaches.

Like iHMC, the CL method involves systematic uncertainties associated with the discretization of spacetime, which are controllable. While no analytic continuation is involved, it should be stressed that the CL method remains a method ``under construction" in the sense that its mathematical underpinning is still under development. However, we interpret the remarkable agreement between CL and iHMC for mass-imbalanced systems, and between CL and a renormalization-group approach across a wide range of interaction strengths (including repulsive couplings), as strong evidence that our CL approach works for the systems studied here, see our discussion below.

\section{Results and Discussion}\label{sec:res}

In this section we present our fully nonperturbative results for the ground-state energy of interacting fermions of unequal masses. Wherever possible, we
compare our results to those obtained by other methods. Additionally, we show the equation of state for the ground-state energy as a function of (attractive
and repulsive) interaction strength across a wide range of mass imbalances. To our knowledge, {this is the first
	determination of the full equation of state for
mass-imbalanced fermions interacting via a contact interaction in one dimension.}

In the following, all values will be shown as dimensionless quantities relative to the ground-state energy of the noninteracting system {\it in the continuum}
$\efg = \f{1}{3}N\ef$ at the same density and particle number.
{We set the number of 1D spatial lattice sites to $N_x = 40$, which we found to be sufficient for the methodological purpose of the
	present work, see also Ref.~\cite{Rammelmueller2017} for a study of the $N_x$ scaling behavior of 1D mass-balanced Fermi gases.}
{The spatial lattice spacing is fixed to unity, which sets the length and momentum scales in our problem.}
The temporal lattice spacing was chosen to be $\tau = 0.05$ and is sufficient to study the
interaction strengths under {consideration~\cite{Rammelmueller2015}. Furthermore, we}
numerically extrapolate to the limit of large imaginary time $\bef$ by fitting a
constant to a few values obtained at sufficiently large propagation times
(following Ref.~\cite{Rammelmueller2015}) to save numerical effort. To carry out that extrapolation, we performed calculations on temporal lattices
as large as {$N_\tau \sim 1500$}, which we found in previous work to be sufficient for the particle
numbers and couplings {considered here~\cite{Rammelmueller2015}.}
Each data point shown was computed using an average of $5000$ decorrelated samples (both in the iHMC and CL approaches).

\subsection{Imaginary mass imbalance}
In order to study the interacting Fermi gas, it is instructive to first investigate its noninteracting counterpart. To calculate the noninteracting energy
on the lattice, we simply sum the single-particle energies and as a function of the mass imbalance $\mbar$ we obtain
\beq
E_{\mbar} = E_{0}\left[\f{1}{1-\mbar^2}\right] = E_{0}\left[\f{1}{1+(-i\mbar)^2}\right],
\label{Eq:noninteractingMassImbalance}
\eeq
where $E_0$ is the corresponding noninteracting energy for mass-balanced systems {\it on the lattice}. {This expression is symmetric in $\mbar$,
	as it should be, since we investigate equally populated spin species. Note also that the energy as obtained from a calculation in the mean-field approximation
	exhibits the same dependence on~$\bar{m}$ as detailed here for the noninteracting system (see, e.g., Ref.~\cite{Braun2015}).}

\begin{figure}[t]
	\includegraphics[width=\columnwidth]{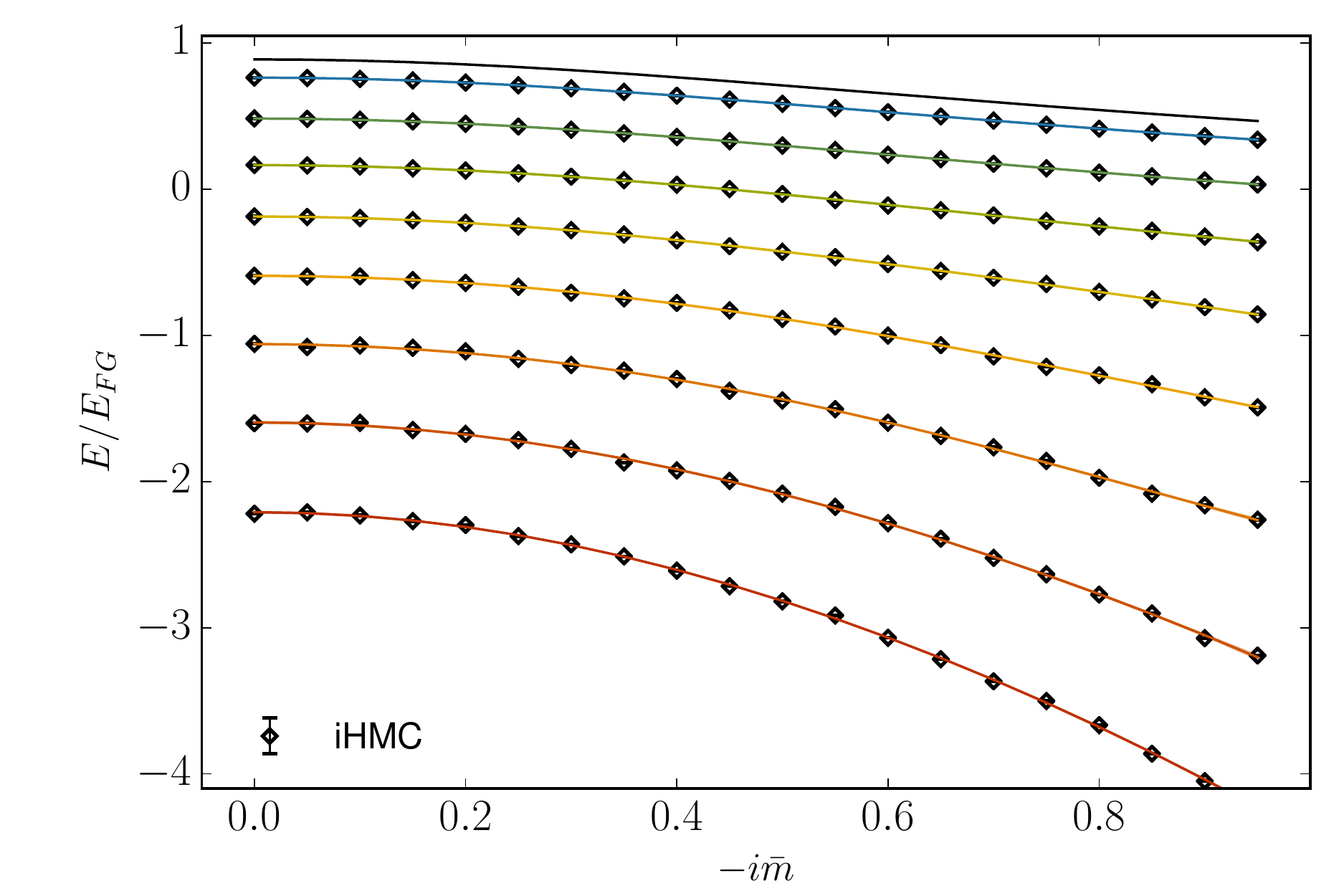}
	\includegraphics[width=\columnwidth]{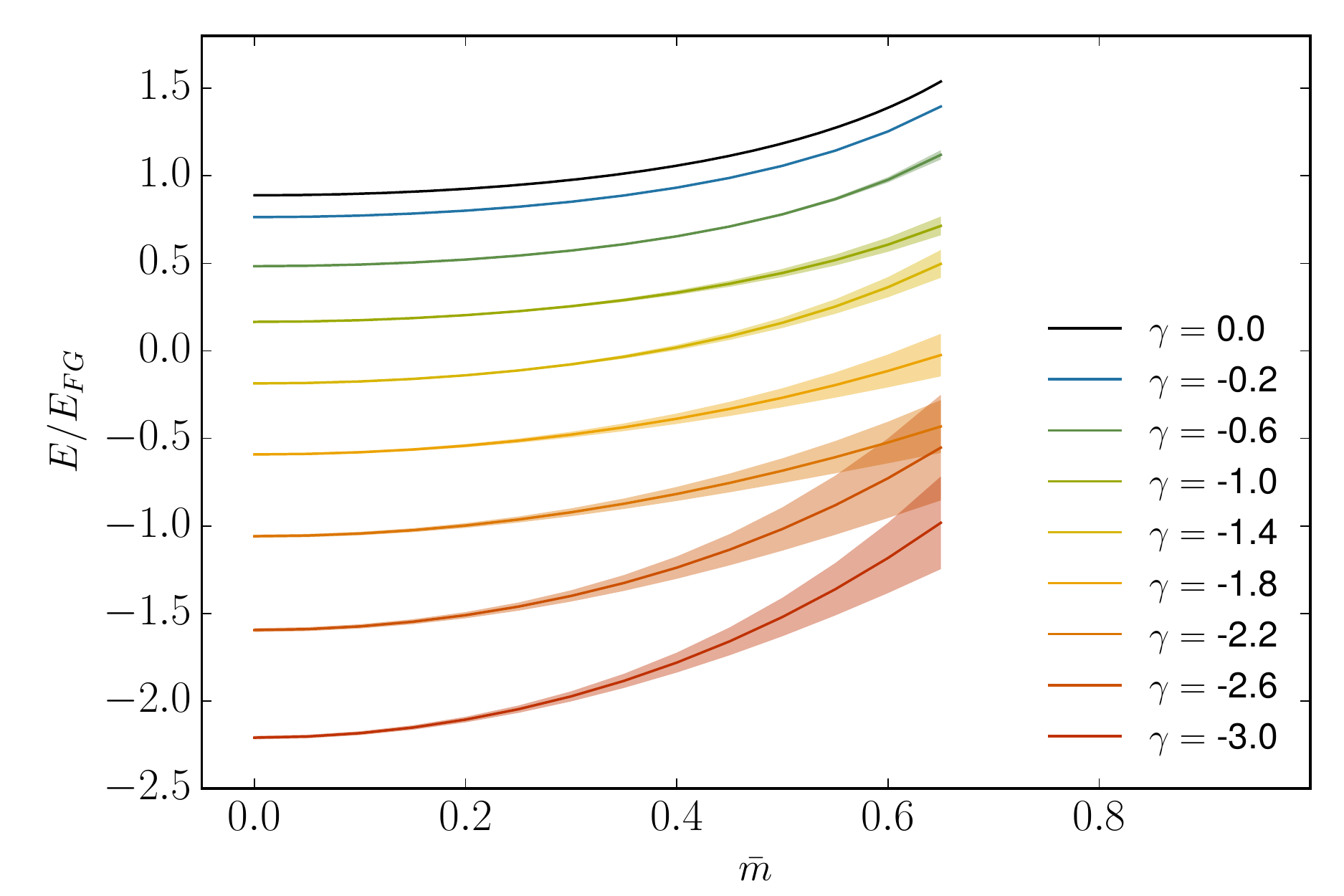}
	\caption{\label{Fig:EOSiHMC} Ground-state energy of $N = 3+3$ fermions as a function of imaginary (top) and real
		(bottom) mass imbalance for various couplings $\gamma$ from weak to strong attractive interaction (lines ordered from top to bottom). Top:
		iHMC results for {imaginary $\bar m$ (black diamonds, statistical error bars are of the size of the symbols) with}
		Pad\'e approximations according to Eq.~\eqref{Eq:Pade} (solid colored lines). The black solid line shows the noninteracting result on the lattice.
		Bottom: analytically continued ground-state energies as a function of real mass imbalance (solid lines). Although the fits as a
		function of imaginary mass imbalance are precise, small uncertainties result in wide confidence bands (shaded areas) when displayed as a function of
		real mass imbalance. The plot range in the bottom panel was limited to $\mbar = 0.65$ due to large uncertainties beyond that point.}
\end{figure}

The top panel of Fig.~\ref{Fig:EOSiHMC} shows our results for the ground-state energy (black diamonds) for various couplings as function of imaginary $\mbar$
along with the noninteracting line (solid black) according to Eq.~\eqref{Eq:noninteractingMassImbalance}.
The noninteracting form suggests the use of a Pad\'e approximant fit to the data, which takes the form
\beq
f(\bar{m}) = \f{\sum_{i\geq 1}{b_i \bar{m}^{2i}}}{1 + \sum_{j\geq 1}{c_j \bar{m}^{2j}}},
\label{Eq:Pade}
\eeq
where the even powers reflect the symmetry under \mbox{$\mbar \to -\mbar$}, and the~$b_j$'s and~$c_j$'s are fit parameters.
The colored lines in Fig.~\ref{Fig:EOSiHMC} represent a least-squares fit of the above form with a polynomial of order~$2\ (4)$ in the numerator (denominator). The
nearly perfect agreement with the numerical data is crucial when performing an analytic continuation to real $\bar m$ as small variations in the fit parameters
can greatly influence the final {results for real mass imbalances.
In principle, higher orders can be included in the polynomials; however, we have found that doing this limits the stability
of the fit procedure. Therefore, we only use the before-mentioned order of the Pad\'{e} approximant in this~work.}

\begin{figure}[t]
	\includegraphics[width=\columnwidth]{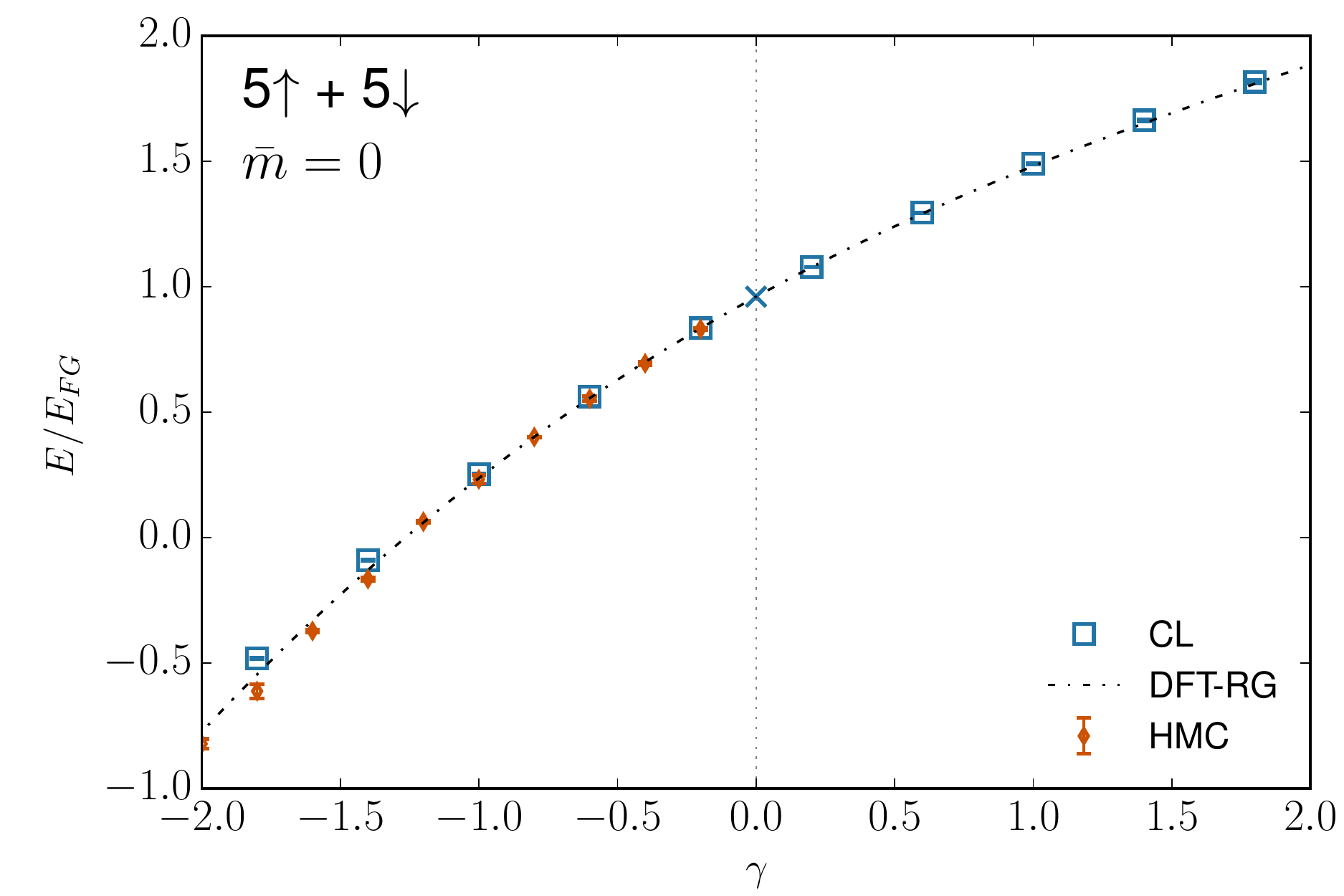}
	\caption{\label{Fig:BalancedCompare} Ground-state energy of $N=5+5$ fermions of equal mass ($\mbar = 0$) as a function of interaction strength computed with iHMC (red error bars), CL (blue squares), and DFT-RG (dash-dotted line).}
\end{figure}

To obtain results for real $\bar m$, we perform an analytic continuation to the real axis via
\beq
i\dm \rightarrow \dm
\label{Eq:wick}
\eeq
which implies
\beq
\mbar \to \f{\dm}{2 m _0}.
\eeq
The results of the analytic continuation are shown in the bottom panel of
Fig.~\ref{Fig:EOSiHMC} along with the 95\% confidence level (shaded). We find
very good agreement with the form of the noninteracting result (solid black line) and the results for the energies are very stable with the order of the Pad\'e
approximant up to $\mbar \sim 0.5\dots 0.6$. For mass imbalances beyond $\mbar \sim 0.6$, however, the associated uncertainties grow rapidly and a
quantitative prediction for the ground-state energy (or any other observable) is not guaranteed, particularly at strong couplings. At very high {imbalances}
(not shown in this plot), it is even possible that the qualitative trend as a function of~$\bar{m}$
changes due to the effect of the higher-order terms in the functional form of the fit. A
possible solution to this issue {may be to use a larger amount of data and a finer grid for the $(-i \mbar)$-axis.
While this is feasible (albeit tedious) in 1D, the numerical effort in 2D and 3D would be definitely prohibitive.}

\subsection{Complex Langevin approach}

{In this section we present our results obtained with the CL approach, as introduced above, and
	benchmark our results with those from our iHMC study and (semi) analytic calculations.
	In what follows,
	we discuss results for the strength of the
	regulating term to be $\xi = 0.1$ and the {target CL {integration step $h_0 = 0.01$.
		From an analysis of the dependence of our results on these parameters, we} found that these values are well suited to
		study the ground-state energy within a precision on the $1\%$ level}.
	We stress, however, that those values could change when considering different quantities
such as correlation functions and density matrices.}

\subsubsection{Mass balanced case: Arbitrary interaction}

We begin by considering the mass-balanced scenario, which provides a valuable cross-check as such systems are accessible with a variety of other
methods. {In particular, we compare our results to those previously obtained with HMC in Ref.~\cite{Rammelmueller2015} for attractive systems
	which have also been found to agree with exact results from the Bethe ansatz.
	Additionally,
	{we show results from a renormalization-group approach to density functional theory based
		on the microscopic interactions defining our model~\cite{Kemler2017Diss}. We}
	will abbreviate this approach as DFT-RG which was put
	forward in Refs.~\cite{Polonyi2002, Schwenk2004, Kemler2013, *Kemler2017}.
	As shown in Fig.~\ref{Fig:BalancedCompare}, we find outstanding agreement among all methods for $-2\lesssim \gamma < 0$ (attractive coupling).
	Moreover, the results from our CL study and those from
	the DFT-RG approach also agree very well for the repulsive case in the regime $0<\gamma \lesssim 2$, where our QMC approach is bound to fail due
	to the sign problem.
	Note that~$|\gamma| \lesssim 2$ is roughly the range
where the DFT-RG approach is able to formulate reliable predictions based on state-of-the-art truncations presently restricted to mass-balanced~systems.}

\begin{figure}[t]
	\includegraphics[width=\columnwidth]{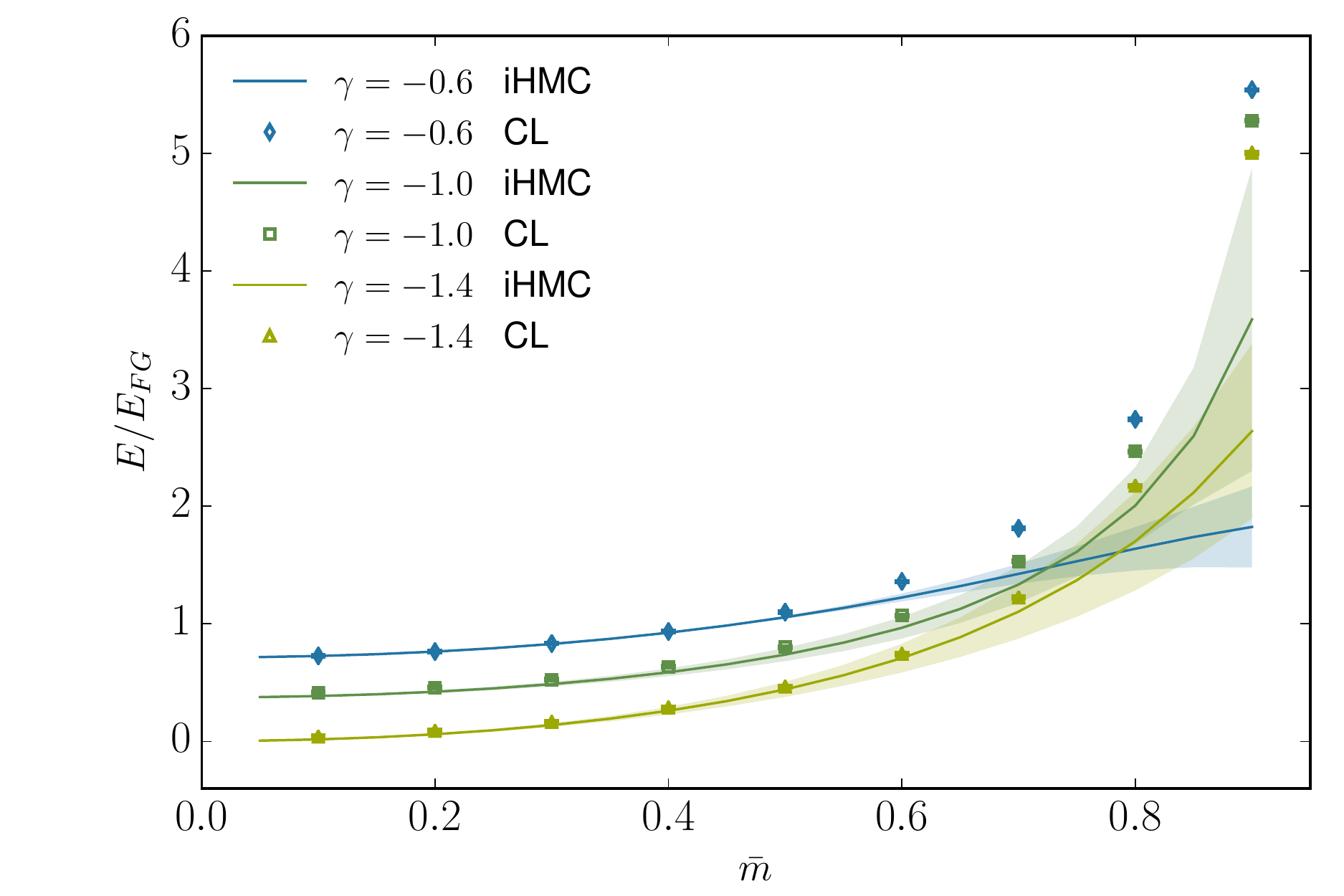}
	\caption{\label{Fig:EOScompare} Ground-state energy of $N=5+5$ fermions computed using iHMC (solid lines) and
		CL (symbols). The shaded areas represent the $95\%$-confidence interval of iHMC data, the uncertainties in the CL data are smaller than the
		symbol sizes. We find agreement in the ground-state energies at low imbalances up to $\bar m \sim 0.5$. Beyond that point the higher-order terms
		in the Pad\'e approximants reduce the curvature of the iHMC lines. The energies calculated with CL increase monotonically as $\bar m$ approaches
		unity {as naively expected for} a species with diverging kinetic energy (zero mass).}
\end{figure}
%

\subsubsection{Mass imbalanced case: Comparison with iHMC}

Motivated by the excellent agreement between CL and other methods in mass-balanced systems, we expand our investigation to mass-imbalanced systems
using the CL approach. As mentioned above, there is no need for analytic continuation, which saves in computational effort since we only have to compute
single data points (as opposed to a grid of data points which is then fitted). Although it is possible to
run calculations for an arbitrary configuration of the fermion
masses $m_\uparrow$ and $m_\downarrow$, we stick to the definition Eq.~\eqref{Eq:idm} introduced with the iHMC method to facilitate a straightforward comparison.

Our CL results are shown in Fig.~\ref{Fig:EOScompare} for various attractive coupling strengths on top of results from iHMC calculations for the same parameter
values. We find excellent agreement between the methods up to $\mbar \sim 0.5-0.6$, which is where the iHMC algorithm incurs large uncertainties
(as mentioned in a previous section). Remarkably, the results obtained with the CL algorithm continue to be smooth well beyond that regime and the
statistical uncertainties are of roughly constant magnitude across all imbalances considered.

\begin{figure}[t]
	\includegraphics[width=\columnwidth]{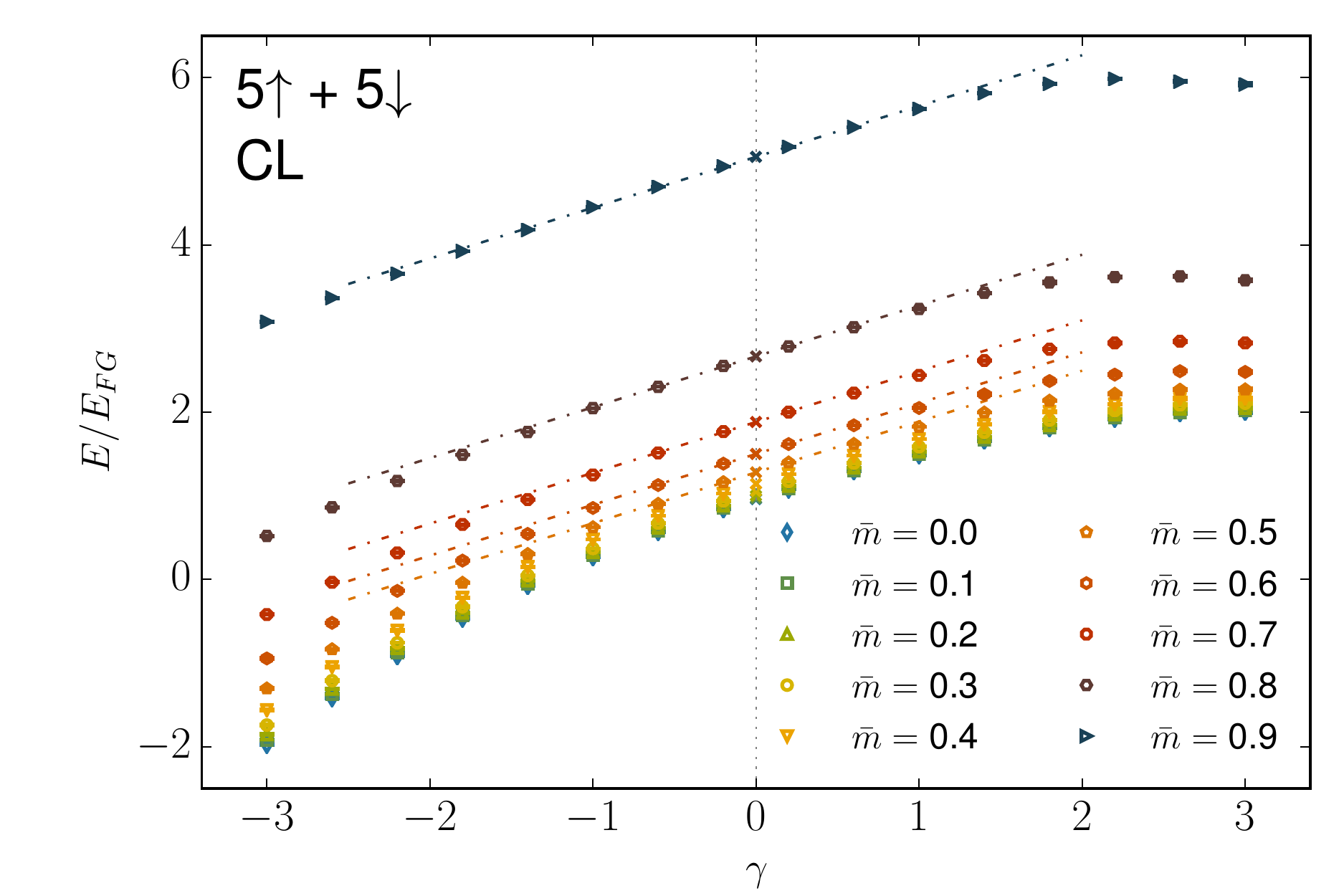}
	\caption{\label{Fig:EOS} Ground-state energy of $N=5+5$ fermions as a
		function of the dimensionless coupling $\gamma$ for several mass imbalances $\mbar$ {as obtained
			from our CL approach. Error bars are of the
			size of the symbols and below.
			The dashed-dotted lines show the first-order perturbative result of Eq.~(\ref{Eq:EOSPT}).}
	}
\end{figure}
%

\subsubsection{Equation of state for arbitrary mass imbalance}

Thus far, we have compared our CL results to various methods and found excellent agreement for all cases considered. Most of parameter space, however, is
generally difficult to access due to analytic and numerical problems, as pointed out above. The CL method is, however, able to
predict values for arbitrary $\mbar$ and across a wide range of both attractive {\it and} repulsive interaction
strengths, although the results for strong repulsion ($\gamma \gtrsim 1$) have to be taken care with some care
at present as we discuss in Appendix~\ref{App:B}.
To underscore this ability, we
present in Fig.~\ref{Fig:EOS} our determination of the equation of state for mass-imbalanced fermions. As can be appreciated in the figure, the results are
smooth as a function of interaction strength and mass imbalance and intersect the correct noninteracting results on the vertical line at $\gamma = 0$.

{It is also evident in Fig.~\ref{Fig:EOS} that the equation of state becomes linear in a region around $\gamma=0$.
	This linear region can be compared with a first-order perturbative calculation of the ground state energy, as shown in
	Fig.~\ref{Fig:EOS}, which
	is given by
	\beq
	\frac{E}{\efg} = \frac{E_{\mbar}}{\efg} + \gamma \frac{24}{\pi^2}\frac{N_{\uparrow} N_{\downarrow}}{(N_{\uparrow} + N_{\downarrow})^2}
	+{\mathcal O}(\gamma^2)\,.
	\label{Eq:EOSPT}
	\eeq
	Here, $E_{\mbar}$ is as in Eq.~(\ref{Eq:noninteractingMassImbalance}). Note that the first-order correction in $\gamma$ does not depend on the
	mass imbalance~$\mbar$, which is reflected in Fig.~\ref{Fig:EOS} in the fact that the slope at $\gamma=0$ does not change as $\mbar$ is increased.
	{Moreover, we observe that} our numerical data agrees very well with this perturbative result around~$\gamma=0$, indicating that our CL approach indeed
	works reliably, at least in the weak-coupling limit. Interestingly, we deduce from this comparison that the
	size of the linear region depends on~$\bar{m}$ and the sign of the coupling~$\gamma$. In fact, the linear region is not symmetric around $\gamma=0$
	and even appears to increase with increasing~$\bar{m}$ for attractive couplings ($\gamma <0$).}

{Our results make the versatility of our CL approach evident. In fact, very promisingly, this
	enables us to predict values for the ground-state energy at couplings and mass imbalances
	relevant to experimental setups where analytic expressions are not available and stochastic calculations have only been of limited
	use so far because of the sign problem. Such experiments
	include for instance mixtures of the fermionic atoms $^{6}$Li and $^{40}$K corresponding to $\mbar \approx 0.74$ but also
	mixtures  with smaller values of~$\bar{m}$ set up from a variety of suitably chosen different fermion species
	(such as ${}^6$Li, ${}^{40}$K, ${}^{161}$Dy, ${}^{163}$Dy, and ${}^{167}$Er)
	in the  future (see, e.g., Refs.~\cite{GrimmPC, Lu2012, Frisch2013}).}

{Finally, it is worth noting a peculiar feature in the equation of state: depending on the actual
	value of the mass imbalance, the energy flattens with increasing coupling constant (repulsive case) and its derivative
	with respect to $\gamma$ appears to vanish in the limit of infinite repulsion. Moreover, the onset of the flattening behavior
	is shifted to larger coupling when the mass imbalance is decreased. This can be seen explicitly in the results for $5+5$ fermions
	in Fig.~\ref{Fig:EOS} but we also observe such a behavior for systems with $3+3$ and $4+4$ fermions. This does not come unexpected
	as our results should only exhibit an explicit dependence on the dimensionless coupling~$\gamma$ and~$\bar{m}$ in the infinite-volume limit but no
	dependence on the actual particle number and box size (i.e. the actual density of the system), provided that
	the box has been chosen sufficiently large.
In any case, this flattening behavior is reminiscent of what is}
sometimes called fermionization, referring to the fact that an interacting system of distinguishable fermions
becomes equivalent to a system of noninteracting identical fermions in the limit of infinite repulsion,
see, e.g., Refs.~\cite{Girardeau1960, Petrov2000, Girardeau2010} for a discussion of this feature for mass-balanced systems.
Evidence
for this behavior has also been observed in experiments~\cite{Kinoshita2004, Kinoshita2005, Haller2009, Zuern2012}.
While this behavior may naively seem like a mere curious feature, it may actually have many interesting consequences. For example, the derivative of the energy
with respect to the coupling is {related to Tan's contact~\cite{Tan2008, *Tan2008a, *Tan2008b, Barth2011}, which}
in turn governs the short-range and high-frequency tails
of all correlation functions.
However, a detailed analysis of the energy and the correlation functions in this truly nonperturbative regime is
beyond the scope of the present work aiming at methodological developments and cross-verification of stochastic methods.
A quantitative study of phenomenologically highly appealing effects, such as the observed flattening behavior of the equation of state, is therefore
deferred to future work as it requires a detailed study of finite-size
effects and the related approach of the numerical data to the thermodynamic limit. Most prominently, it requires a
detailed analysis of systematic effects associated with the CL approach in case of strong repulsive couplings~$\gamma\gtrsim 1$, see also
Appendix~\ref{App:B}.
Still, detailed quantitative studies of at least the onset of this flattening behavior for mass-imbalanced system
appears now in reach based on the present developments.

\section{Summary and Outlook}
We have computed ground-state properties of 1D Fermi gases by means of two stochastic numerical methods, namely iHMC and CL. Both methods are able
to produce fully nonperturbative results. While the iHMC approach performs well for low to intermediate mass imbalances, large mass imbalances remain
elusive due to the instability of the analytic continuation. {Remarkably, the CL method possesses no such restriction and is
	capable of producing quantitative results across all mass imbalances. For small mass imbalances, also accessible to our iHMC approach, the corresponding
results agree very well.}
Although this technique has been known for more than three decades, applications are
remarkably scarce in nonrelativistic scenarios; our work aims to fill that gap. Moreover, we have shown excellent agreement with other methods
wherever possible, including with perturbative results at small coupling $\gamma$.

A word of caution may be in order at this point. It is known that the CL method, being still an approach under construction, may converge to
an incorrect answer (see e.g. Ref.~\cite{Aarts2010a}). It is also possible for the analytic continuation of iHMC results to the real axis to yield an answer that
depends strongly on the choice of fit function. However, the fact that both nonperturbative methods yield essentially the same result in wide swath of
parameter space (i.e. as a function of both $\gamma$ and $\bar m$) is remarkable and supports the idea that the answer
is indeed correct.

Finally, we have shown the full equation of state as a function of interaction strength and mass imbalance. To the best of our knowledge, there is
no previous determination of the equation of state of 1D mass-imbalanced fermions {over such a wide range of mass imbalances and coupling strengths.
	Thus, although more detailed studies of systematic effects are required to push our calculations towards high precision, our main result of
	Fig.~\ref{Fig:EOS} can already be considered as a first prediction for
	future ultracold atom experiments with Fermi mixtures.} {In particular, we have found a feature in the equation of state which appears particularly
	pronounced at intermediate and large mass imbalances where the equation of starts to flatten and approaches a $\bar{m}$-dependent constant
	already at comparatively small values of the coupling,~$\gamma \sim {\mathcal O}(3)$,
	which appears to point to corresponding significant changes in the short-range (or high-frequency) behavior of correlation functions. A detailed analysis
of this strong-coupling regime is deferred to future work.}

Our use of periodic boundary conditions aims to reproduce the uniform system,
which {has now been realized in experimental setups with {{flat-bottom traps~\cite{Hueck2017}} in two-dimensional systems}.
	However, our calculations can be straightforwardly
	extended to harmonically trapped systems, hard-wall confinement, as well as higher dimensions (see, e.g., Refs.~\cite{Berger2014, Berger2015, Braun2015}).
The} discussion of these systems,
however, is left to subsequent studies.

\acknowledgments
{
	This work was supported by HIC for FAIR within the LOEWE program of the State of Hesse and the U.S.
	National Science Foundation under Grant No. PHY{1452635} (Computational Physics Program).
	Numerical calculations have partially been performed at the LOEWE-CSC Frankfurt.
	The authors gratefully acknowledge many helpful discussions with A. G. Volosniev.
}

\appendix
\section{{Effects of the regulator in\\ the complex Langevin approach}}\label{App:A}

\begin{figure}[t]
	\includegraphics[width=\columnwidth]{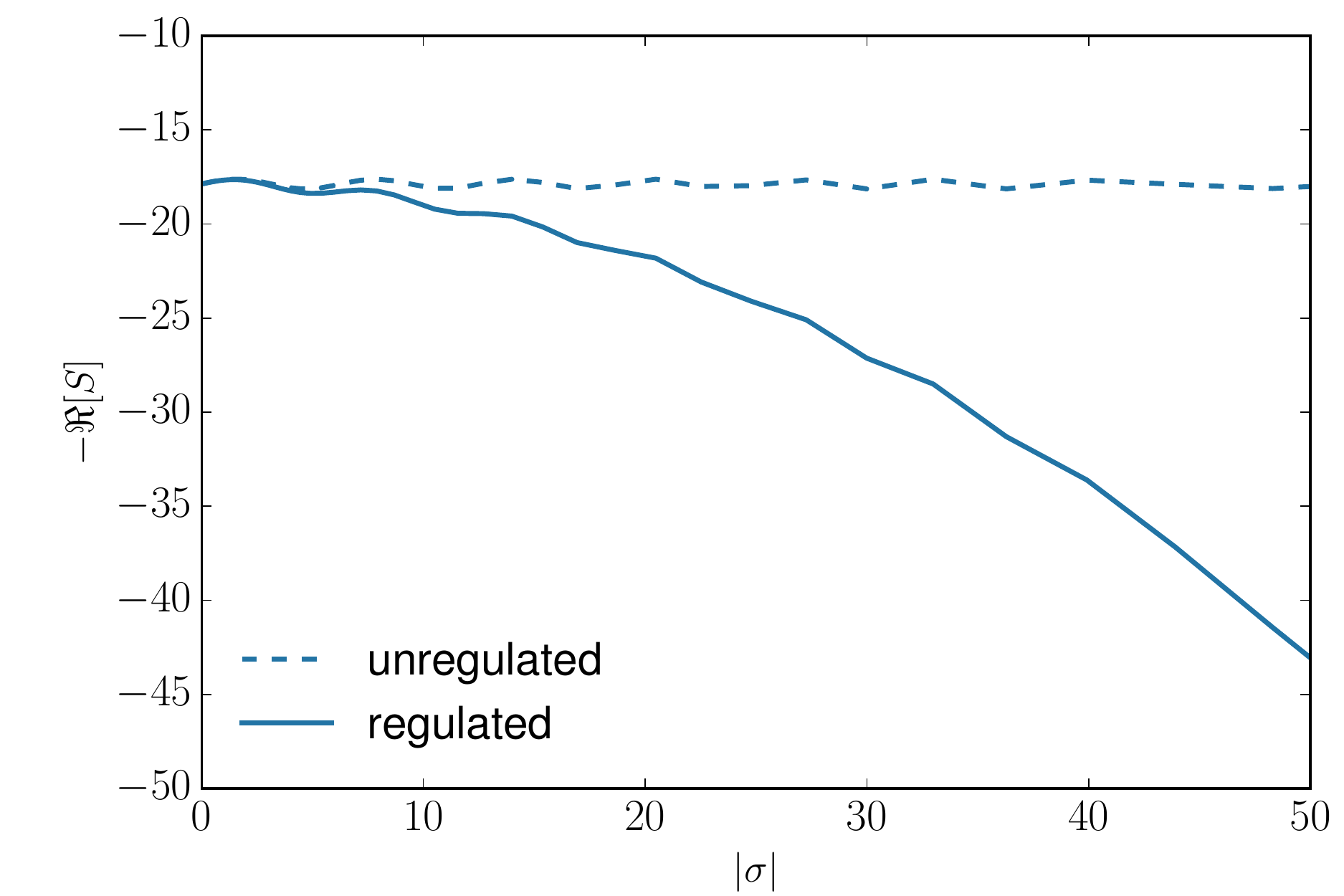}
	\caption{\label{Fig:CLLargeSigma} Real part of the
	logarithm of the fermion determinant (see Sec.~\ref{sec:mbm} for our conventions)
	as a function of the
	magnitude of the auxiliary field (at a randomly chosen point on the lattice)
	for an exemplaric choice of parameters (random seed, time of CL evolution, and coupling)
	for an unregulated (i.e.~$\xi=0$; dashed lines) and an action regulated by choosing~$\xi=0.1$ (solid lines).}
\end{figure}

\begin{figure*}[t]
	\begin{tabular}{@{}cc@{}}
		\includegraphics[width=\columnwidth]{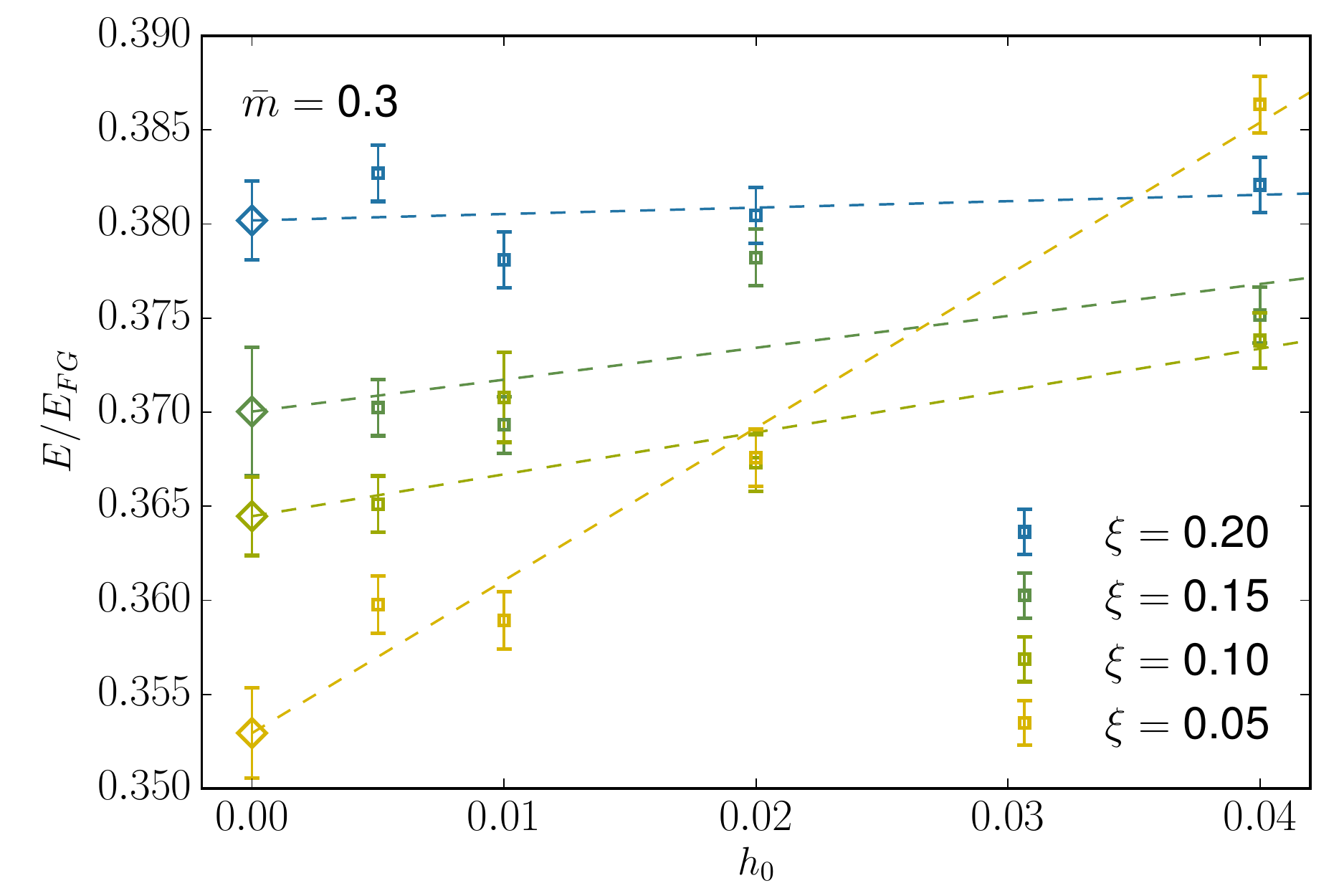}  &
		\includegraphics[width=\columnwidth]{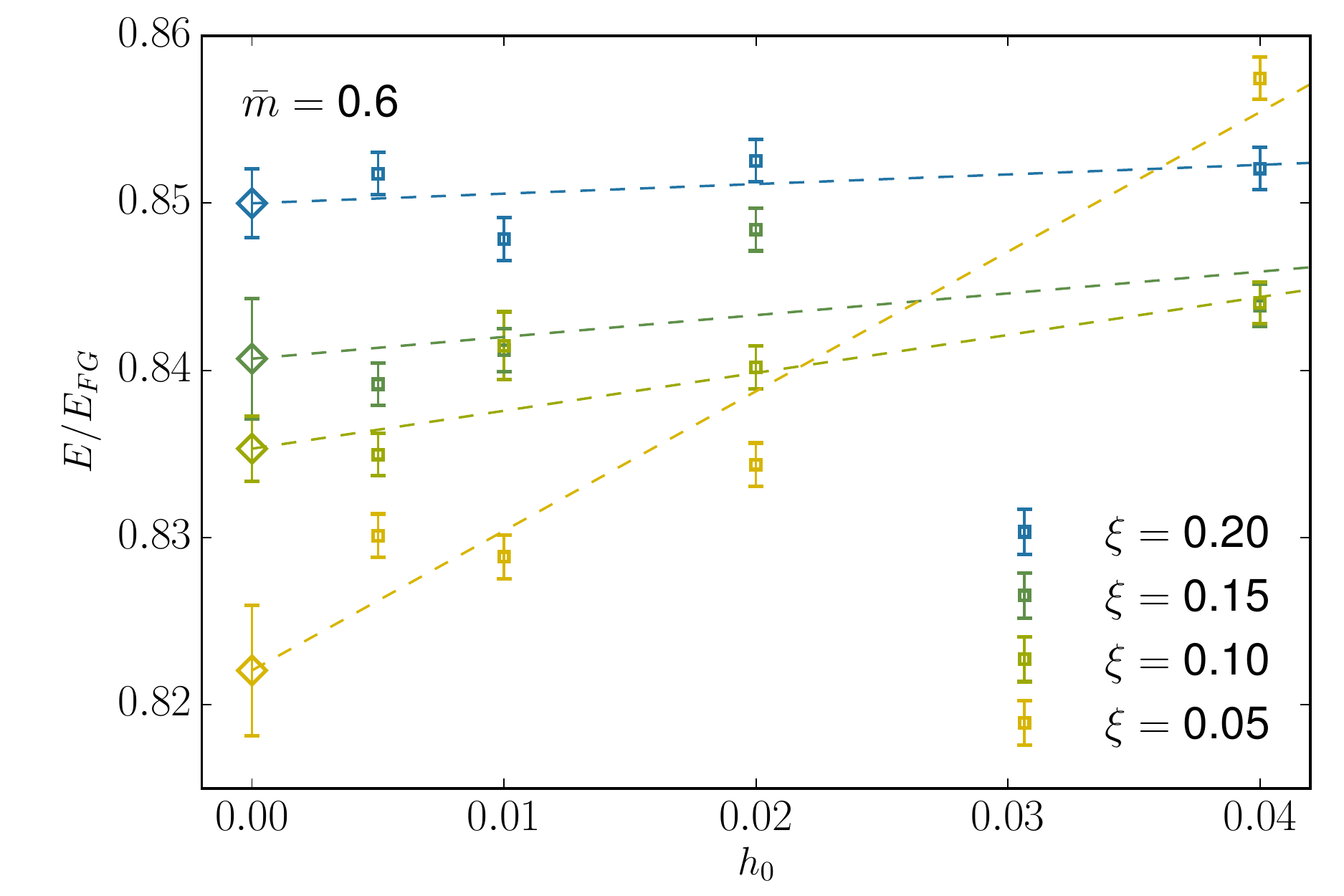} \\
		\includegraphics[width=\columnwidth]{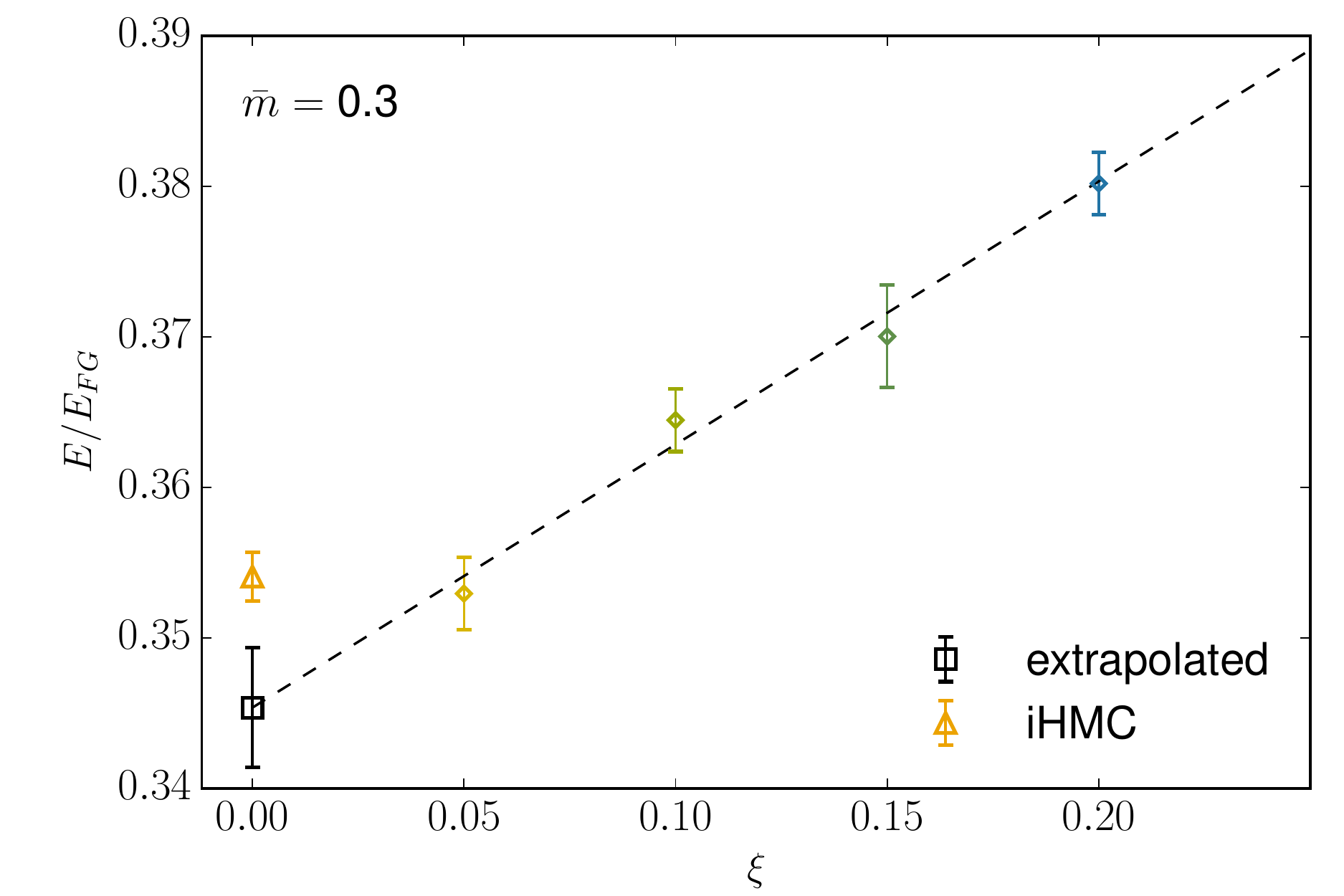} &
		\includegraphics[width=\columnwidth]{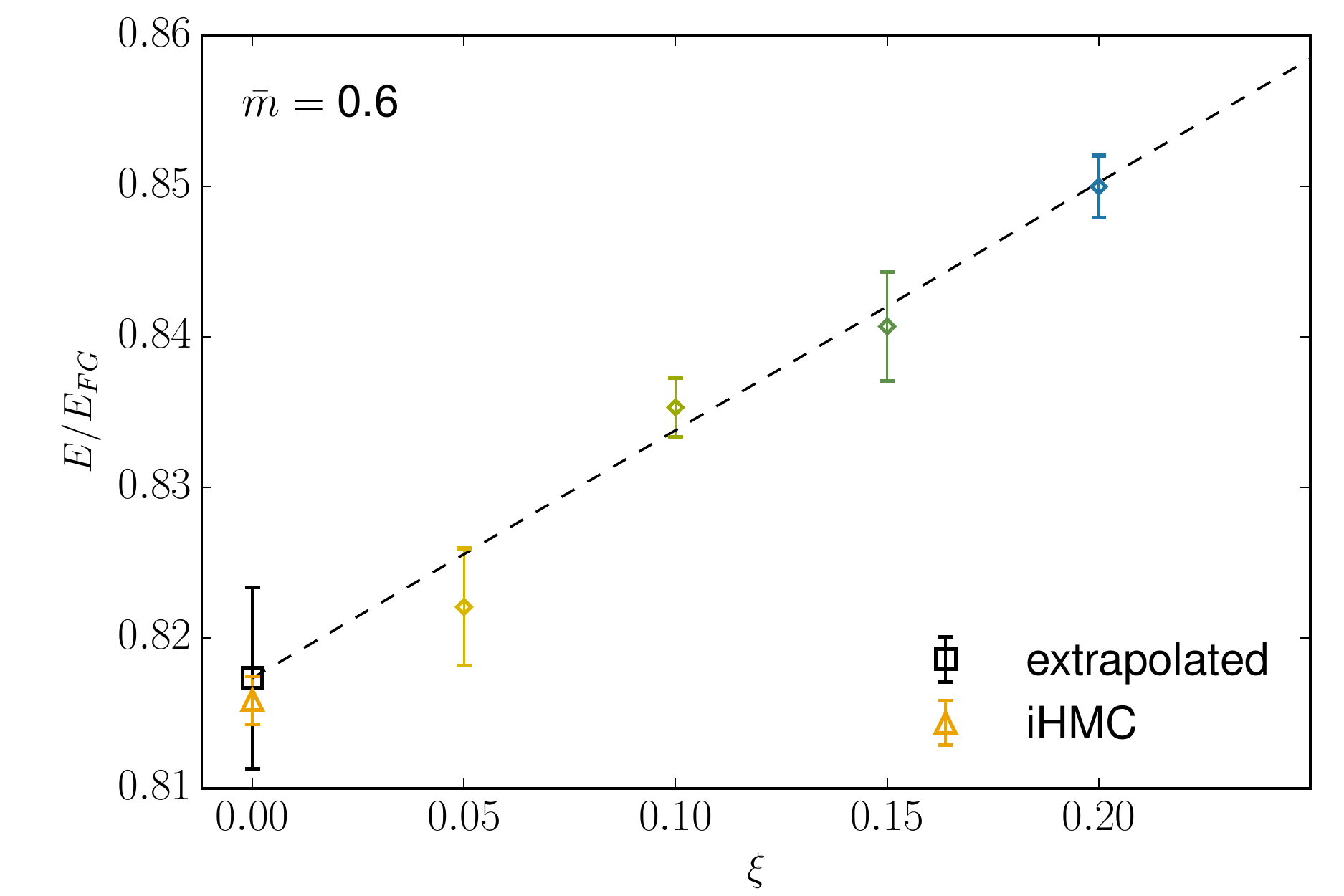} \\
	\end{tabular}
	\caption{\label{Fig:CL_extrapolations} {Top panels: Ground-state energy of {a system with} $N=5+5$ fermions
		{with mass imbalance $\mbar = 0.3$ (top left panel) and $\mbar =0.6$ (top right panel)} as a function
		of the target CL step $h_0$ for different strengths $\xi$ of the regulating term {appearing
			in the CL equations}. The interaction strength is set to $\gamma = -1.0$ in both cases.
		Dashed lines represent linear fits {which have been used} to extrapolate to the limit $h_0 \rightarrow 0$. The latter are
		marked by diamonds. Bottom panels: Ground-state energy of
		a system with $N=5+5$ fermions
		with mass imbalance $\mbar = 0.3$ (bottom left panel) and~$\mbar =0.6$ (bottom right panel)
		as obtained
		from an extrapolation to~$h_0=0$ is shown as a function of the regulator strength~$\xi$.
		The interaction strength is set to $\gamma = -1.0$ in both cases. Dashed lines represent linear fits of the data. For comparison,
		we also show the results obtained from our iHMC approach via analytic continuation.
		{Note that, whereas the error bars on the CL data points
			originate from statistical errors, the error bars on all extrapolated CL values as well as the iHMC values
	refer to errors from associated fits.}}}
\end{figure*}

In this Appendix we investigate the dependence of our results on two numerical parameters which
appear in the CL equations~\eqref{Eq:discreteCL}: the regulator strength~$\xi$
and the step size~$h_0$ entering the solver for the CL equations.

{In Ref.~\cite{Aarts2017}, it has been shown that the probability measure has to decay sufficiently fast in the
limit of large fields~$\sigma$ in order to render the associated CL study reliable.
The insertion of a regulator term in our study is related to this issue. Indeed,
the regulator term is included to control the excursions in the imaginary
direction of the field $\sigma$. In Fig.~\ref{Fig:CLLargeSigma}, we show~$-S[\sigma]$ as a function of~$|\sigma|$
in order to show the result on the action~$S[\sigma]$ of
scaling up in magnitude the value of $\sigma$ at a particular location (starting from an otherwise typical field
configuration in the CL evolution) which
illustrates the necessity of the regulator. Specifically, we}
show $-\Re [S[\sigma]]$, i.e. the real part of the logarithm of the fermion determinant
in the absence of the regulator as well as the corresponding answer with the extra term $-\xi \sum_{x} \sigma^2(x)$ for $\xi = 0.1$. As can be appreciated in the figure,
without the regulating term the action remains at best approximately {constant. Even worse, we also find cases where the action grows as}
the magnitude of $\sigma$ increases. It is for these reasons that the $\xi$ term is needed.

We now turn our attention to studying the influence of the target step size $h_0$ and its interplay
with the regulator term $\xi$, and illustrate the extrapolation procedure for two values of the mass-imbalance paramter~$\bar{m}$.

As mentioned in the {main part of this work}, within the CL approach,
it is {useful} to employ an adaptive step size $h_t$ for the integration of the equations of
motion \cite{Aarts2010}.  {In our approach, the step size is}
determined by rescaling the target step size $h_0$ with the maximum value of the sigma-drift $D_{\tn{max}}$ on the spacetime 
	{lattice $\Lambda$:
\beq
D_{\tn{max}} = \max_{i\in\Lambda} \left|\left.\frac{\delta S[\sigma]}{\delta \sigma}\right|_{\sigma=\sigma_i}\ -\ 2\xi\sigma_{i}\right|^{2}.
\label{Eq:driftCL}
\eeq
Naturally, this} renders the results dependent on the target step size and
appropriate extrapolations to vanishing {$h_0$ are required}. In the top panels of Fig.~\ref{Fig:CL_extrapolations},
the dependence on~$h_0$ is shown for {systems with $N=5+5$ particles} with
$\mbar = 0.3$ and $\bar{m}=0.6$ on a spatial lattice {with} $N_x = 40$ sites. Furthermore, data sets for multiple
values of the {regulator strength $\xi$} are shown. {In order to enable a comparison
with iHMC results, we consider an attractive coupling. To be specific, we have set the coupling
to $\gamma = -1.0$ here, but we add that the general behavior of the CL results for systems
with a repulsive interaction is the same. For our extrapolations to vanishing target step size in this work, we
have always performed a linear
fit of the data which appears to be justified given our data sets, see top panels of Fig.~\ref{Fig:CL_extrapolations}
for an illustration. Note that the} slope decreases with an increasing regulator strength.

Once the extrapolation to $h_0=0$ has been performed for a given system, the dependence of the
$h_0$-extrapolated values on the regulator strength $\xi$ has to be considered. For the latter, we observe
an approximately linear behavior. Therefore, we use again a linear fit to extrapolate $\xi\rightarrow0$.
{This} is shown in the lower panels of Fig.~\ref{Fig:CL_extrapolations}. 
Remarkably, the extrapolated values agree very nicely with those obtained {by our iHMC approach via
analytic continuation.

We conclude by noting} that the results from these {two extrapolations are essentially
independent of the order in which they are performed, i.e. if we first} extrapolate to~$\xi=0$
and {then perform the extrapolation to~$h_0=0$, we obtain the same results for the energy within
the presented error bars.

\section{Distribution of the path-integral measure in the complex Langevin approach}\label{App:B}

\begin{figure*}[t]
	\begin{tabular}{@{}cc@{}}
		\includegraphics[width=\columnwidth]{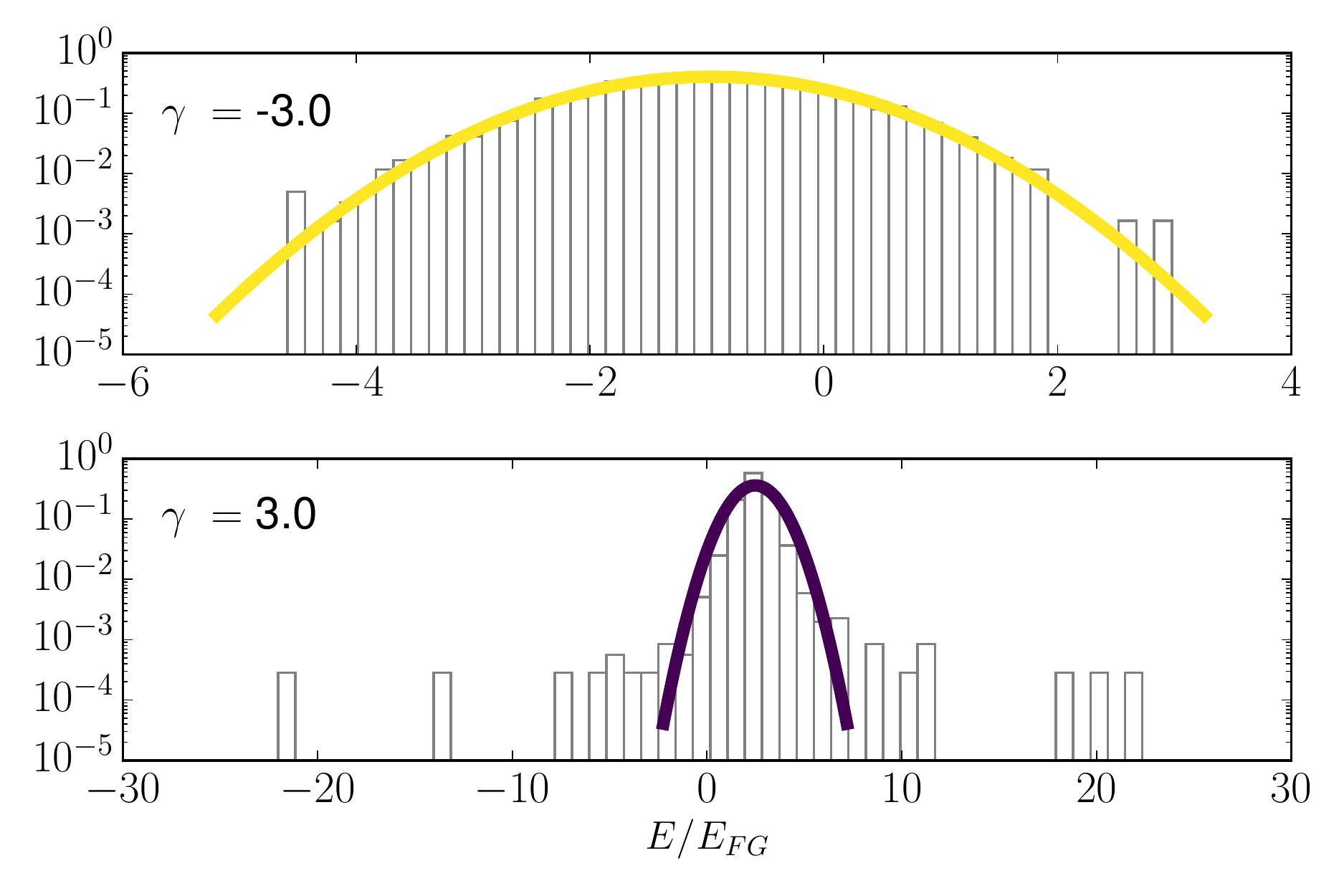} &
		\includegraphics[width=\columnwidth]{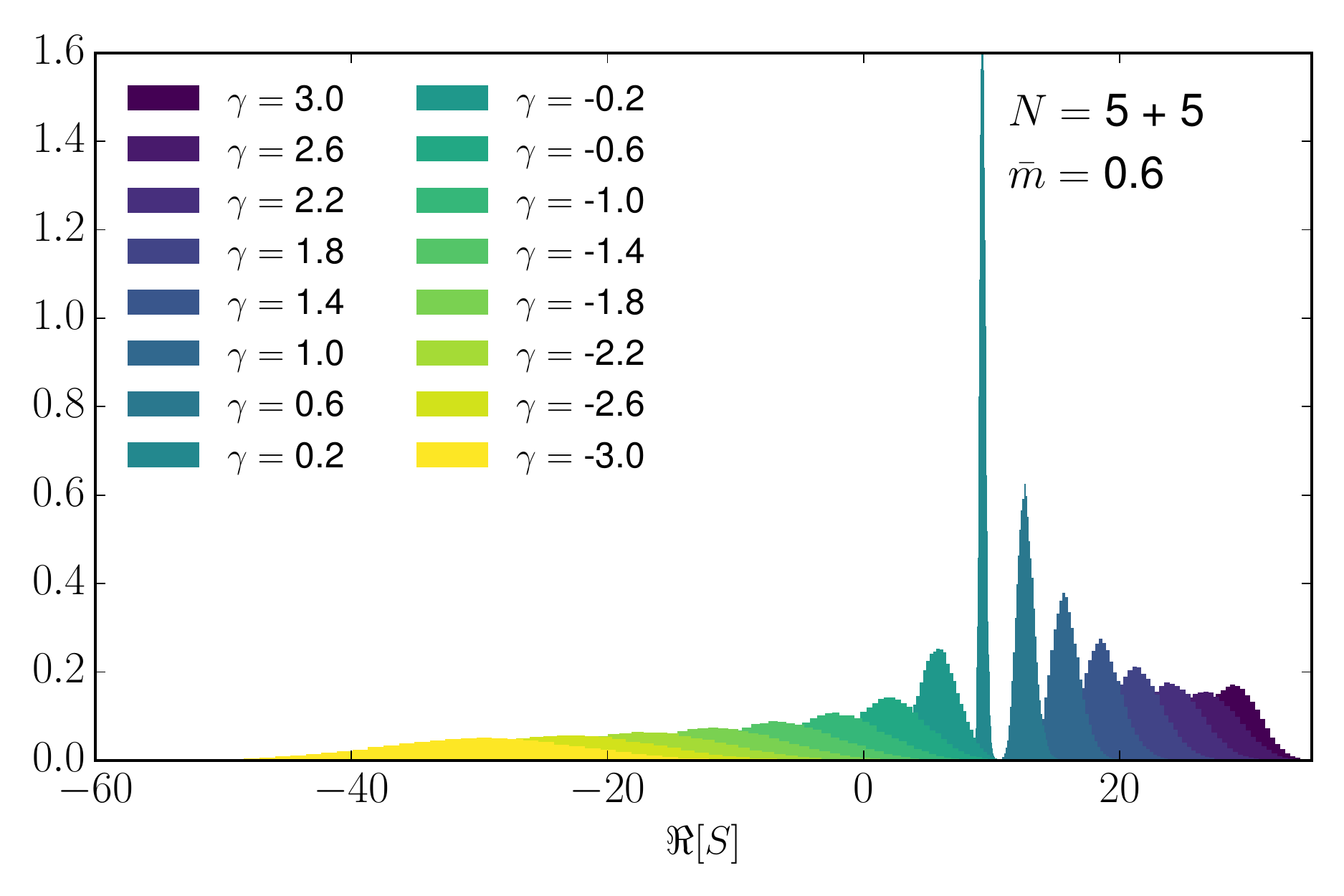} \\
	\end{tabular}
	\caption{\label{Fig:CL_measures} Left panels: (logarithmic $y$-scale)
	of the measured ground-state energies for an attractively (top) and a repulsively (bottom) interacting system. While the former follows a
	Gaussian distribution, the latter exhibits so-called ``fat tails" as a consequence of a {\it signal-to-noise} problem.
	Right panel: distributions of the real part of the path integral measure, i.e. the real part of the negative logarithm of the
	fermion determinant for various systems from strongly attractive to strongly repulsive (from left to right).}
\end{figure*}

In order to ensure correct convergence of observables calculated with our CL approach, we
investigate the distribution of ground-state energies as well as the path integral measure in this appendix.
To this end, we consider a system composed
of $N = 5+5$ particles with a mass imbalance of $\mbar = 0.6$. The general conclusions, however, are
valid also for all other systems studied in this work.

During the evaluation of the results obtained from stochastic methods, such as iHMC and CL, it is instrumental to monitor
histograms of the calculated observables in order to gain an insight into
the behavior of the simulations. Furthermore, it is common practice to define the error bar as the standard deviation over all samples with an assumed
Gaussian distribution. A deviation from such a distribution may hint to systematic errors. In the left panel of Fig.~\ref{Fig:CL_measures},
we show the distribution of the ground-state energies for a strongly attractive (top) and strongly repulsive (bottom) case.
While the histograms associated with attractive systems follow a Gaussian very closely, the repulsive systems exhibit so-called
``fat tails", i.e. an excessive amount of ``outliers"
with respect to the assumed normal distribution. In fact, more generally speaking, we find that the distributions in the latter case
do not exhibit a fixed variance. The origin of this problem is depicted in the right panel of Fig.~\ref{Fig:CL_measures}, where we show the distribution (as obtained
from the CL time evolution)
of the real part of the action $S$: Depending on the absolute value of the coupling,
the distribution peaks at small positive to large negative values for attractive systems (from weak to strong attraction). For increasing repulsion, on the other hand, the peak wanders to large positive values and
the imaginary part of the action is found to be a flat distribution whereas the imaginary part is strongly localized about zero in the attractive case.
Note that an increase of the value of the action corresponds to a decrease of the probability measure~${\rm e}^{-S}$.
Eventually, we even find that the associated probabilities will decrease below the machine precision~($\sim 10^{-16}$). This unavoidably implies that information is lost
and immediately leads to a poor \emph{signal-to-noise ratio}. As a consequence, error bars calculated with the assumption of a Gaussian distribution become unreliable.

The occurrence of signal-to-noise problem is not limited to the CL approach and has been studied for other methods based on a
Hubbard-Stratonovich transformation~\cite{Hao2016} (see also Refs.~\cite{DrutPorter1,DrutPorter2}, where a very similar signal-to-noise problem was solved for the calculation of entanglement entropies).
Although not implemented in our approach, methods have been proposed to mitigate this issue. Within the CL approach, however, zeroes of the
determinant entail~$S\to\infty$ and therefore exhibit an additional problem, namely singularities in the drift term. These singularities, when encountered,
possibly spoil the computed expectation values~\cite{Aarts2017}. This may be rated as conceptual problem, in contrast to
a vanishing signal due to excessive noise. The latter is at least in principle solvable by (drastically) increasing the sample number. However, note that it may very well be
that both problems are related in our case. In fact, a zero of the determinant yields a vanishing probability which is reminiscent of the situation of strong repulsion
as reported above. At present, we cannot resolve whether the CL dynamics at strong repulsive couplings is governed by zeroes of the determinant or
whether we only deal with a conventional {\it signal-to-noise} problem which could at least in principle be solved.

In conclusion, we have found that our simulations yield ``fat-tailed distributions" of observables in the repulsive
regime (at least for strong repulsion), which are associated with potentially spoilt
expectation values. For small to intermediate repulsion~($0< \gamma\lesssim 1$) the problem appears to be absent or at least strongly suppressed and
our results agree very well with DFT-RG, perturbation theory, and exact solutions
from the Bethe ansatz~\cite{2006JPhA...39.1073O} where applicable (i.e. for~$\bar{m}=0$).
Finally, we have found that our CL approach behaves very well for attractive systems for the studied range
of mass imbalances which is not accessible for conventional HMC approaches and exact calculations
with the Bethe ansatz.

\bibliography{refs_1dmib}

\end{document}